\documentclass[conference]{IEEEtran}
\usepackage{times}
\usepackage{latexsym}
\usepackage{url}
\usepackage[english]{babel}
\usepackage[utf8]{inputenc}
\usepackage{mathrsfs,amsmath}
\usepackage{amsfonts}
\usepackage[colorinlistoftodos]{todonotes}
\usepackage{multirow}
\usepackage{amssymb}
\usepackage{hhline}
\usepackage{enumitem}
\usepackage{dblfloatfix}
\usepackage{subcaption}
\usepackage{breqn}
\usepackage{placeins}
\usepackage{array}
\usepackage{booktabs}
\usepackage{listings}
\usepackage[]{scrextend}
\usepackage[]{upgreek}
\usepackage{color,soulutf8}

\usepackage{relsize,etoolbox}
\AtBeginEnvironment{quote}{\smaller}

\usepackage{microtype}

\begin{document}
\title{Using Sentiment Information for Preemptive Detection of Toxic Comments in Online Conversations}

\author{\IEEEauthorblockN{Éloi Brassard-Gourdeau}
\IEEEauthorblockA{\textit{Department of computer science and software engineering} \\
\textit{Université Laval}\\
Québec, Canada \\
eloi.brassard-gourdeau.1@ulaval.ca}
\and
\IEEEauthorblockN{Richard Khoury}
\IEEEauthorblockA{\textit{Department of computer science and software engineering} \\
\textit{Université Laval}\\
Québec, Canada \\
richard.khoury@ift.ulaval.ca}
}

\maketitle
\begin{abstract}
The challenge of automatic detection of toxic comments online has been the subject of a lot of research recently, but the focus has been mostly on detecting it in individual messages after they have been posted. Some authors have tried to predict if a conversation will derail into toxicity using the features of the first few messages \cite{conversations}. In this paper, we combine that approach with previous work on toxicity detection using sentiment information \cite{subversive}, and show how the sentiments expressed in the first messages of a conversation can help predict upcoming toxicity. Our results show that adding sentiment features does help improve the accuracy of toxicity prediction, and also allow us to make important observations on the general task of preemptive toxicity detection.
\end{abstract}

\section{Introduction}
\label{sec:introduction}
Billions of messages are sent online every day and hidden among them are millions of toxic and harmful messages. For instance, studies have shown that 17\% of internet users receive cyber-bullying messages (a type of toxic messages), with a disproportionate number of targets being women (19\%), low-income people (24\%), and homosexuals (34\%) \cite{hango2016cyberbullying}, and that 10\% of people develop depressive or suicidal thoughts as a result of these messages \cite{ADL2019}. This makes the development of accurate and efficient content moderation systems a high priority. Most of the studies so far have focused on single-line detection. In other words, the models developed look at each message individually and decide whether it should be classified as toxic or not. While such systems can be good at detecting toxic messages once they are written \cite{meanbirds,deeper,vanhee2018}, they cannot predict whether upcoming messages in a conversation will feature toxic content or not. This ability to flag interactions for moderation before they turn toxic would be hugely beneficial both for community moderators, allowing them to intervene more quickly and efficiently, and for users, preventing them from being targeted by toxic messages in the first place. This is the goal of the task of preemptive detection. It is however impossible to do when considering only a single message in isolation, and requires a model of the entire conversation.

In \cite{conversations}, the authors study the pragmatic devices used early in a conversation and their usefulness for preemptive detection. In this paper, we build upon their work by adding sentiment information \cite{subversive} to their model. Our intuition is that the sentiments expressed early in a conversation can help predict more accurately if it will degrade into toxicity later on or not. If true, this would run counter to the observations of \cite{conversations}, which found sentiment information to have no predictive power in this task. The rest of this paper is structured as follows. After a review of the relevant literature in section \ref{sec:background}, we will quickly go over the sentiment detection tool in Section \ref{sec:Sentiment}. In the same section we will also study how our sentiment features can be added to the system of \cite{conversations} and present the resulting preemptive detection tool. We will conduct an in-depth analysis of our results when using the dataset of \cite{conversations} in Section \ref{sec:results}. To expand on this study, we then perform a second set of experiments on another dataset in Section \ref{sec:other_context}. Finally, we draw conclusions on preemptive detectio and the use of sentiment information in Section \ref{sec:Conclusion}.

\section{Related Work}
\label{sec:background}

The challenge of toxic content detection in online conversations has been studied since 2012. Various topics have been covered, such as hate speech detection \cite{warner2012detecting,nobata2016abusive} and cyberbullying detection \cite{reynolds2011using,meanbirds,deepcyber2018}, and many architectures have been adapted and trained successfully for this task, including SVMs \cite{warner2012detecting,vanhee2018}, logistic regressions  \cite{nobata2016abusive}, and neural networks \cite{deepcyber2018}. However, even the most recent work only focuses on single-line detection, meaning determining whether a comment that has already been posted is toxic or not by itself and outside the context of the conversation where it appears.

One of the first and only studies on toxicity prediction at the conversation level is that of \cite{conversations}. The authors showed that certain features in the first messages of a conversation, such as the use of first or second person pronouns and the presence of certain politeness strategies, can help predict if that conversation will remain healthy or if it will degrade and lead to toxic messages later on. Their work inspired the authors of \cite{preemptive}, who also worked on preemptive moderation. They trained and tested an SVM using TFIDF-weighted unigrams and bigrams as well as a BiLSTM using their own word embeddings. They were ultimately dissatisfied with their results, however the fact they focused only on words and didn't use more sophisticated features such as those in \cite{conversations} may be the cause. Finally, the authors of \cite{liu2018} did hostility presence and intensity prediction on Instagram comment threads using a variety of features, ranging from n-grams and word vectors to user activity and lexicons. The features are used to train a logistic regression model with L2 regularization. The authors conclude that there are four main predictors for hostility: the post author's history of receiving hostile comments, the presence of user-directed profanity in the thread, the number of distinct users posting comments in that thread, and the amount of hostility so far in a conversation. However, none of these studies examined the impact of sentiment information in preemptive detection, which will be the main focus of our paper.

\section{Conversation Model}
\label{sec:Sentiment}
\subsection{Sentiment Detection Tool}
\label{sub:sentimenttool}
The authors of \cite{subversive} implemented a sentiment detection system in order to study whether sentiment information can help detect toxic content in a subversive setting (where users deliberately misspell toxic words to mask them from keyword filters). They found that sentiment information did correlate to toxicity, and could be used to improve the accuracy of toxic message detection systems, both in a normal and in a subversive setting.

The sentiment detection tool implemented in that paper, which we will reuse in this one, is heavily inspired by previous works such as \cite{ohana2012case,nielsen2011new,tumsare2014opinion}, where the authors used sentiment lexicons, such as SentiWordNet or General Inquirer, to detect the sentiment of a message. Our tool combines three popular lexicons, namely SentiWordNet\footnote{\url{http://sentiwordnet.isti.cnr.it/}}, Afinn\footnote{\url{https://github.com/fnielsen/afinn}} and Bing Liu\footnote{\url{https://www.cs.uic.edu/~liub/FBS/sentiment-analysis.html}}. The authors found previously that these three lexicons have different strengths and weaknesses, and thus complement each other well. SentiWordNet is the biggest lexicon and assigns a positive and negative score between 0 and 1 to each word. Afinn assigns a single score between -5 and 5; scores under zero meaning the words are negative. The Bing Liu lexicon has a positive and a negative word list. The lexicons are combined by splitting each into lists of positive and negative words for each of four parts-of-speech (noun, verb, adverb, and adjective), and normalizing the sentiment scores between 0 and 1. 

The sentiment detection tool begins by detecting sentiment-carrying idioms in the messages. For example, while the words "give" and "up" can both be neutral or positive, the idiom "give up" has a clear negative sentiment. Several of these idioms can be found in our lexicons, especially SentiWordNet (slightly over $60,000$). When detected, these idioms are marked so that our algorithm will handle them as single words. Next, it uses the NLTK \textit{wordpunkt\_tokenizer} to split messages into words, and the \textit{pos\_tagger} to get the part-of-speech of each word. Each word is then assigned a positive and a negative score, which is the sum of the score it has in the positive and negative lists of each of the three lexicons. A message is represented by the score of its three most positive words and its three most negative words. This gives us a total of 6 sentiment features for each message. For more details as to why the tool is built this way, please refer to \cite{subversive}.

\subsection{Model and features}
\label{sec:Features}
The authors of \cite{conversations} split their conversation pragmatic features into two categories: 13 politeness strategies and 6 rhetorical prompts. The first category focuses on the use of politeness, such as greetings, gratitude, or the use of "please", and of impoliteness, such as direct and strong disagreement or personal attacks. The second category captures six domain-specific conversation prompts, which are six clusters of conversations discovered by an unsupervised technique trained on a different dataset that includes similar types of discussions. A new message's distance to each of these six clusters gives the six prompt features. This gives a total of 19 features per message, and the authors compute them for the first two messages of a given conversation, thus getting a set of 38 features. Using these features, the authors train a logistic regression model to predict if a conversation will derail into toxicity based on its first two messages. The authors have made their code available publicly\footnote{https://github.com/CornellNLP/Cornell-Conversational-Analysis-Toolkit}. More details on these features and the regression model built from them can be found in the original article. 

Our version of the model builds upon theirs by adding the 6 sentiment features measured by the sentiment tool for the same two messages. We also computed another sentiment feature representing the overall tone of the first two messages. This feature is computed by taking the sum of positive word scores of the first message and subtracting the sum of negative word scores to determine if the message is overall positive or negative, doing the same for the second message, and determining if the conversation starts with two positive messages, a positive followed by a negative, a negative followed by a positive, or two negative messages. This information is encoded as a one-hot vector of length 4. In total, there are thus 38 text features from \cite{conversations} and 16 sentiment features we added, for a total of 54 conversation features.

\begin{figure*}[b]
    \centering
    \includegraphics[width=0.48\textwidth]{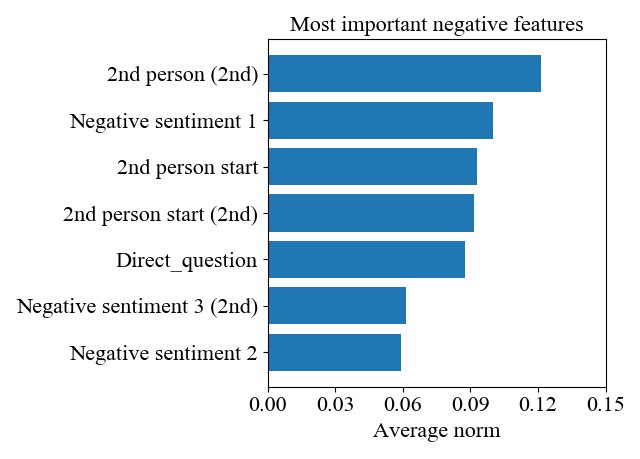}
\includegraphics[width=0.48\textwidth]{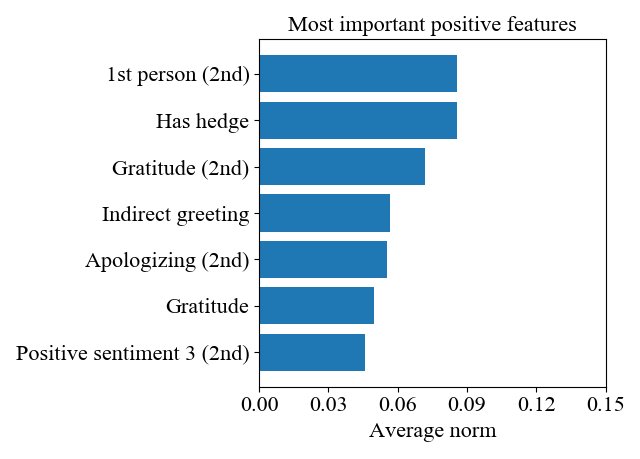}
    \caption{Feature importance when using 3 positive sentiment features and 3 negative sentiment features. The "(2nd)" refers to the feature on the second message, while its omission refers to the first message.}
    \label{fig:feature1}
\end{figure*}

\subsection{Data and Training}
\label{sec:Data and Training}

The dataset created for \cite{conversations}, which is available publicly along with their code, is a set of user conversations taken from the edit pages of English Wikipedia. The authors used Persepctive API\footnote{https://www.perspectiveapi.com/} to pre-filter the conversations and keep only the ones with potentially toxic content. They further filtered to keep those conversations that started in a civil way, meaning that didn't have any toxic content in the first two messages. Moreover, they required conversation pairs, one derailing and one staying civil, from each Wikipedia page. This resulted in 1,270 conversation pairs from 582 different pages with an average length of 4.6 messages.

Our model is trained using \textit{Scikit-learn}'s \textit{LogisticRegression} and \textit{SelectPercentile}, with a grid search on hyperparameters \textit{C} between $10^{-4}$ and $10^4$ and \textit{percentile} between $10$ and $100$\footnote{\textit{C} representing the regularization and \textit{percentile} representing the percent of features to use.}. Training was done using a 5-fold cross validation. Apart from increasing the number of folds from 3 to 5 for more consistency between runs, all the training parameters are exactly the same as the ones in \cite{conversations}.

\begin{figure*}[h]
    \centering
    \includegraphics[width=0.48\textwidth]{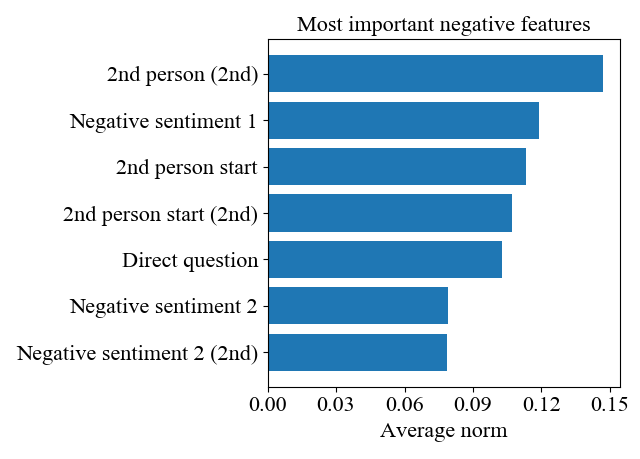}
\includegraphics[width=0.48\textwidth]{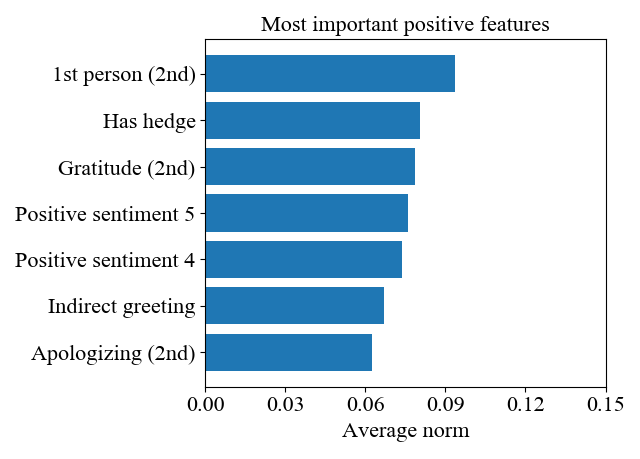}
    \caption{Feature importance when using 5 positive sentiment features and 2 negative sentiment features. }
    \label{fig:feature2}
\end{figure*}

\section{Results and Analysis}
\label{sec:results}
As in \cite{conversations}, our experiments consist in taking a pair of conversations, looking at their first two messages, and predicting which of the two conversations will remain healthy and which one will derail into toxicity. All the results presented in the following section are the average of 10 separate runs, where we randomized the data split.

\subsection{Sentiment Features}
Our first experiment considers the predictive accuracy of sentiment information alone. In fact, the authors of \cite{conversations} did include the sentiment lexicon of \cite{Liu:2005} in their research, and used it to extract two sentiment features per message. Their features were "has negative" and "has positive", each being 1 if a negative or positive word from the lexicon was present in the message and 0 otherwise. However, after testing these features, they concluded that sentiment was barely better than random chance at predicting toxicity, and they didn't include them in their set of 38 text features.

The goal of our first experiment is thus to validate that the sentiment features are in fact predictors of upcoming toxic messages. We trained and tested the model using four setups: using the original sentiment features of \cite{conversations}, the sentiment word features, our tone features, and all sentiment features combined. The results are presented in \ref{tbl:baseline}.

\begin{table}[ht]	
\centering
\begin{tabular}{l|c|c}
\hline
Test & Features & Accuracy \\ \hhline{=|=|=}
Original sentiment & 4 & 51.3 \\
Our sentiment & 12 & 55.7 \\
Our tone & 8 & 50.8  \\
All features & 24 & 55.8  \\
\hline
\end{tabular}
\caption{Prediction accuracy using sentiment features.}
\label{tbl:baseline}
\end{table}

Our results firstly confirm that the minimalist sentiment features of \cite{conversations} are nearly equivalent to a random chance guess. This is likely due to the fact that over 70\% of the messages containing a negative word also have a positive word, making it nearly impossible to discern a toxic message from a healthy one based on that information alone. Likewise, our tone information carries nearly no useful information. However, our more detailed word features do show an interesting predictive ability. Finally, combining all features together gives no gain compared to just using the word features; an unsurprising result, given that the other features seem to contain no predictive information.

This shows that, when it comes to sentiment information, it is not the overall sentiment of a message that is useful, but individual words. That level of detail is  missing from both the original sentiment features (which only indicated whether positive or negative sentiment exist) and our tone features (which only indicate whether positive or negative sentiment is stronger). It is however present in the sentiment word features, which indicates the sentiment of the three most positive and most negative words of each message without making a judgment on whether the message overall is positive or negative. That finer level of granularity seems to be where the predictive information is found.

From this point forward, we will drop the tone features from our model, since they are not predictive of toxicity. This will leave 12 sentiment features and a total of 50 conversation features.

\subsection{All Features}
\label{sec:allfeats}

Our next experiment consists in training and testing our model with and without the sentiment features. The goal is to highlight the gain in prediction accuracy that comes from including sentiment features. The results of that experiment are given in Table \ref{tbl:mainresults}.

\begin{table}[ht]	
\centering
\begin{tabular}{l|c|c}
\hline
Test & Features & Accuracy \\ \hhline{=|=|=}
Text features & 38 & 58.6 \\
Text + sentiment & 50 & 60.5 \\
\hline
\end{tabular}
\caption{Prediction accuracy with and without sentiment features.}
\label{tbl:mainresults}
\end{table}

In all 10 runs of our test, we found that the model including sentiment features consistently performs better than the one without. Our results using text features alone are consistent with those of \cite{conversations}, and adding sentiment features improves the prediction on average by 2\%. This is consistent with the findings in \cite{subversive}, where it was found that sentiment information improved toxicity prediction by 3\%. 

\subsection{Predictive Features}
It is interesting to examine which sentiment features contribute the most information to the prediction of how a conversation will develop. To do this, we take the average norm of the coefficient score of the logistic regression for each of the 50 features over the 10 runs of our experiment. The most informative features are simply those with the highest positive or negative coefficients, while features with coefficients around 0 have no influence on the prediction.

We found that the most predictive features were consistent from run to run. They are listed in Figure \ref{fig:feature1}, along with their average coefficients. The top text features found match those identified in \cite{conversations}. In addition to those, four of the sentiment features are among the 14 most predictive features found by the regression model. 

For predicting conversations that will feature toxic messages, the strength of the first and second most negative words in the first message and of the third most negative word in the second message are all strong predictors. This indicates that strong negative words in both first messages will likely cause the conversation to degrade. Combined with the fact that second-person pronoun use in both messages are also strong toxicity predictors, this may indicate conversations that begin with directed negative sentiments towards other participants.

On the other hand, only one of the sentiment features is among the strongest predictors of whether a conversation will remain healthy. It is the strength of the third most positive word in the second message. This is an interesting difference with the toxic case: while strong negative words are clear predictors of upcoming toxic messages, strong positive words are not predictors of healthy messages, but lower-ranked positive words are. This may indicate that abundance, not strength, of positive sentiment is what matters to predict health.

In order to verify that theory, we re-trained and re-tested our model several times using between 1 and 7 positive or negative sentiment features. The best combination we found was using 5 positive sentiment features and only 2 negative ones, and this 56-feature model offered an improvement of 1\% on prediction accuracy compared to the 50-feature model of Table \ref{tbl:mainresults}. The most predictive features in that test are shown in Figure \ref{fig:feature2}. For toxicity prediction, nothing has changed, save for the fact the third negative word of the second message has disappeared (as the feature is no longer part of the model) and the second negative word of the second message becomes the seventh most predictive feature (it was eighth previously). For health prediction, we can see that the newly-added features of the fourth and fifth positive words of the first message are now among the top predictors, beating out the third positive word from \ref{fig:feature1}. This confirms our earlier intuition.

\subsection{Case Studies}
Figure \ref{fig:quote} has an example of the first two messages of a conversation that was mispredicted as healthy using the text features alone, but was correctly predicted as leading to toxic messages by the classifier with sentiment features.

\begin{figure}[h]
    \begin{quote}
    (1) I'm sorry to say it, but I'm \underline{pretty} sure this is the only option left. This discussion has been so repetitive it's unbelievable. The mediation cabal has all but ceased, and the mediation of this talk page has \textbf{failed}. The RFC also did not work. I can see no other way to reslve the issue other than ArbCom. What does everyone else think? 
    
    (2) It's a \underline{pretty} \textbf{useless} process.  Mostly Admins listing Hebrew as a language, or displaying Israeli symbols on their user pages will respond they have no  \textbf{problem} with the biased edits.  \textbf{Worst}-case they ban you for suggesting the article needed comment or some type of oversight. 
    \end{quote}
    \caption{First two messages of a derailing conversation, with major good words underlined and major bad words in bold.}
    \label{fig:quote}
\end{figure}

The first message uses the first person and apologizes, both text features that predict a healthy conversation, and no other predictive text features are present in either message. As a result, the text-based system predicts they will lead to a healthy conversation. In reality, this conversation eventually degrades into the users attacking each other with messages such as: "[username] actually blames others", "it's your problem", "you are just trying to find an excuse to take jabs at me" and eventually "[username] shut up".

When taking sentiment information into account, the picture is quite different. Both messages contain only a single strong positive sentiment word, the word "pretty" (score of 0.59). The other positive words are very weak, and the fourth and fifth positive words of the first message are "all" and "think" (scores of 0.04 and 0.02 respectively). On the other hand, the first message has two strong negative words, "failed" (score of 0.47) and "unbelievable" (score of 0.46), and the second message has three even stronger ones, "useless", "problem" and "worst" (scores of 0.67, 0.60, and 0.78 respectively). Negative features dominate these messages, and as a result our model predicts correctly that this conversation will derail into toxicity.

This example highlights one reason why the top positive words are not predictors of health: they can be used as modifiers to enhance negative words, as is the case of the word "pretty" in "pretty useless". We believe another reason the strongest positive words are not good predictors is sarcasm, which uses one or two very strongly positive words to convey a negative message. However, we found no examples of sarcasm in our dataset, so we could not confirm that hypothesis.

\begin{figure}[h]
    \begin{quote}
    (1) not vandilism 

    (2) \underline{well} sorry about replacing bands.but you \textbf{dumb} \textbf{cunt} fireworks is also a punk pop band
    \end{quote}
    \caption{First two messages of a derailing conversation, with major good words underlined and major bad words in bold.}
    \label{fig:quote2}
\end{figure}

The sample conversation of Figure \ref{fig:quote2} is an example of the impact of strong negative words. The second message in particular contains an apology (positive indicator), the strong positive word "well" (score of 0.46), and uses the second person (negative indicator). However, most people will pinpoint the two negative words as the strongest indicators this conversation will degrade. In fact, if those words were removed from the message, it would become a much more civil conversation. This illustrates how one or two strongly negative words can change the tone of a message and the flow of a conversation.

\section{Gaming Chat Moderation}
\label{sec:other_context}

\begin{figure*}
    \centering
    \includegraphics[width=0.48\textwidth]{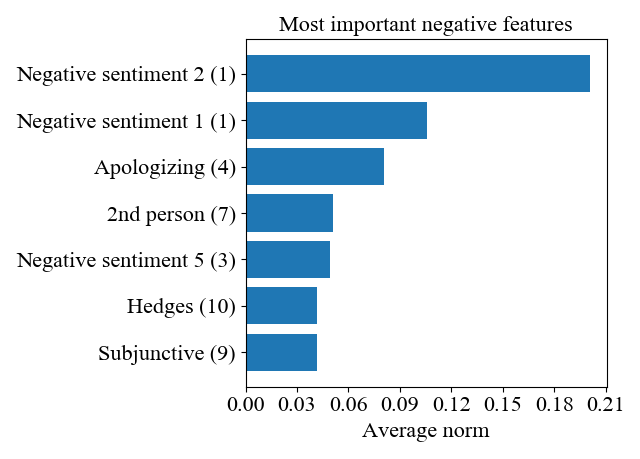}
\includegraphics[width=0.48\textwidth]{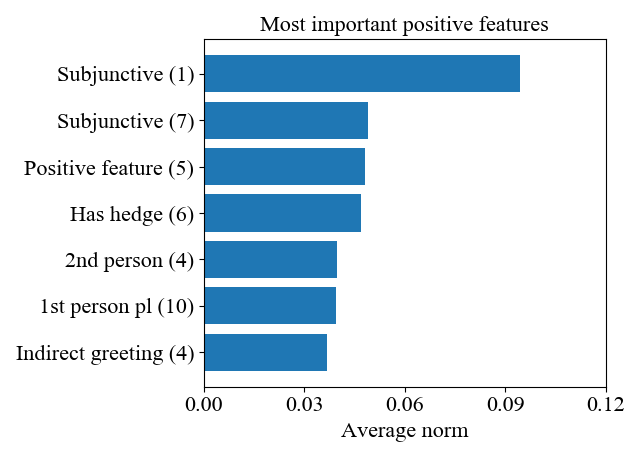}
    \caption{Feature importance in the gaming chat dataset. The number in parenthesis refers to the message's position before the reported message.}
    \label{fig:feature3}
\end{figure*}

To validate the generality of our results, we decided to apply our model to a completely different setting from Wikipedia talk pages: live in-game chat conversations from a popular video game\footnote{The dataset was provided by Two Hat Research Corp. with permission from the gaming company.  The data was pseudonymized and users have agreed to have their chat used for moderation purposes. The data can not be shared publicly due to its sensitive nature.}. This dataset consists of 26,964 different conversations of up to 50 messages, with most messages being very short, around 4 words only. This makes it very different from the Wikipedia dataset, in which conversations are on average less than 5 messages long but messages are on average 58 words long. The last message of each conversation was reported by a user, and then a decision was made by a community moderator to either take action on the reported message or ignore the report. The dataset is balanced, with 54\% of messages moderated and 46\% ignored. 

There are several other significant differences with the Wikipedia dataset. Unlike an edit discussion which has a well-identified initial message, a gaming chat conversation begins when the chat room is created and is continuously ongoing after that, with players joining and leaving at will. The dataset's 50-message conversations are actually composed of the reported message and the previous 49 messages. Moreover, the Wikipedia dataset contains mostly two- to four-person conversations, while very often over a dozen players can chat simultaneously (together or in intertwined separate discussions) and be present in the 50-message conversation. 

The purpose of this experiment is slightly different from the previous one: while we still want to determine if it is possible to predict if a conversation will derail into toxicity (meaning in this case that it will need moderation) from earlier messages, and to measure which text and sentiment features are the strongest predictors of this, we are no longer working with conversation pairs. Consequently, instead of choosing which of two conversations is most likely to go awry, we predict for each conversation individually if it will go awry or not, which is a much harder problem. Moreover, since the first message in a conversation is not the first message of the chatroom, we are not making a prediction from the beginning of a conversation but from an arbitrary point in the middle of it. Finally, taking only the first two messages as before would represent on average 8 words, which is not enough information to make a prediction from. Consequently, we use instead the 10 messages prior to the reported comment to predict whether the unseen final message will be toxic and require moderator action or not. This is thus a true preemptive detection challenge as well as a predictive moderation challenge: based on 10 messages, we are predicting whether an unseen 11th message will be moderated or not.

We will use the same 19 text features and 7 (5 positive and 2 negative) sentiment features per message as before. However, with 10 messages instead of 2, this means our model will have 260 features as input instead of 50. Moreover, we expect that message chronology will be a lot more important in a 10-message sequence than with 2 messages. Consequently, we decided to try two different models. The first one is the same logistic regression model as before. The second model is a recurrent neural network, specifically a uni-directional GRU with a kernel of 40 and a linear layer taking the final state of the GRU and producing a binary output. A recurrent neural network is a natural choice for a problem with a lot of features where chronology is important, and a similar model was used in \cite{subversive} for single-line toxicity detection and found to works well. 

The data was randomly split 70/20/10 into training/validation/testing sets. We once again did 10 training and testing runs, using a different random split each time and 5-fold cross-validation within each run. Average results over all 10 runs are presented in Table \ref{tbl:gamingchatresults}. These results confirm that adding sentiment information helps improve the prediction of toxic conversations. The gain is greater for the logistic regression model, which in fact fails to make a prediction better than random chance without sentiment information. The RNN fares better, probably because it can better handle the large number and sequential nature of the features, but it still gains 1\% by including sentiment information.

\begin{table}[ht]	
\centering
\begin{tabular}{l|c|c|c}
\hline
Model & Features & Accuracy & F1 score\\ \hhline{=|=|=|=}
Regression & 190 & 50.9\% & 0.556 \\
Regression & 260  & 57.4\% & 0.564\\
RNN & 190 & 60.6\% & 0.686 \\
RNN & 260 & 61.7\% & 0.691 \\
\hline
\end{tabular}
\caption{Results for both models using text features alone (190 features) or text and sentiment features (260 features).}
\label{tbl:gamingchatresults}
\end{table}

As before, we use the average coefficient score of each feature over the 10 runs to rank the features by predictive importance. The top features are shown in Figure \ref{fig:feature3}. There are some differences with the results of the Wikipedia test. Most notably, the subjunctive feature\footnote{Expressions such as "would you" and "could you".} is a strong predictor of healthy conversations in this experiment. Looking more closely, this feature is predictive of unmoderated conversations two-thirds of the times it appears; however, it appears in less than 1\% of chat conversations. This difference is therefore not significant in practice.

On the other hand, the coherent aspects with the previous experiment are very interesting. In both experiments, the features 'has hedge'\footnote{'Has hedge' refers to the presence of hedges, or mitigating words, like 'think', 'almost', 'rather', etc. This differs from the feature 'hedges', which looks for dependencies and requires the subject of the message to express this hedge.}, the use of 1st person pronouns, and indirect greetings\footnote{The presence of words like 'hey', 'hello' or 'hi'.}, are indicators of healthy conversations, while strong negative-sentiment words are indicators of an upcoming toxic comment that will need to be moderated. Moreover, unlike with the 'subjunctive' feature, these features all occur in a significant number of the conversation. This confirms that the method is generalizable and can be applied to different types of online conversations.

Next, we considered the question of which messages in the conversation contain the most predictive features. To this end, we considered the 26 (10\%) most predictive positive and negative features, and grouped them per message. The results, given in Table \ref{tbl:msgfeat}, show that features predicting both health and toxicity can be found throughout the conversation. However, while health predictors are distributed evenly in the conversation, toxicity predictors are concentrated in the final three messages. This indicates that a healthy conversation is an ongoing process, but a few bad messages can very quickly turn the tides of the conversation and lead to toxic messages being posted. This also indicates a limit to preemptive moderation: long-term predictions are not valid, and one must focus on clues in the latest messages. To confirm this, we ran the experiment again using only 3 messages before the reported message instead of 10. The results are almost identical to before: the logistic regression classification has an accuracy of 57.7\% with sentiment and 51.7\% without, while the RNN has an accuracy of 61.8\% with sentiment and 60.9\% without. It seems clear, then, that the previous seven messages did not contribute significantly to the prediction accuracy.

\begin{table}[ht]
\centering
\begin{tabular}{l|c|c|c|c|c|c|c|c|c|c}
\hline
Mess & 10 & 9 & 8 & 7 & 6 & 5 & 4 & 3 & 2 & 1 \\
\hhline{=|=|=|=|=|=|=|=|=|=|=}
Pos & 3 & 3 & 3 & 2 & 1 & 2 & 4 & 3 & 3 & 2 \\
Neg & 2 & 2 & 1 & 3 & 2 & 1 & 3 & 6 & 2 & 4\\
\hline
\end{tabular}
\caption{Number of positive and negative predictive features per message before the reported message.}
\label{tbl:msgfeat}
\end{table}


\section{Conclusion}
\label{sec:Conclusion}
In this paper, we studied how sentiment information can be used as a feature for the tasks of preemptive toxicity detection. We conducted this study using the sentiment detection tool developed in previous work \cite{subversive}, the conversation features and logistic classifier of \cite{conversations}, and two very different online conversation datasets. The results of our experiments allow us to draw some important conclusions that can guide both future research and practical implementations of preemptive detection tools:
\begin{enumerate}
  \item Sentiment information is indeed a predictor of toxicity. Using it improves a system's performance by between 1\% and 6\%, which is consistent with previous results in \cite{subversive}. This notably runs counter to previously-published results from other authors that indicated that sentiment information performs no better as a predictor of toxic messages than random chance.
  \item Sentiment information is found at a fine granularity, at the individual word level. Using coarser information, such as overall message sentiment, is not informative. This may explain the above-mentioned contrary previously-published results.
  \item It takes a lot of weak positive words to maintain a healthy conversation, but only a few strong negative words can turn a conversation toxic. 
  \item The features that are predictive of health and toxicity are consistent between very different formats of conversations, and a preemptive detection system is therefore generalizable to multiple different online communities. 
  \item A conversation turns negative very quickly, and consequently negative predictors are concentrated in the few most recent messages. This puts a natural limit to the range of preemptive detection. This range limit seems to be of 3 messages in our results.
\end{enumerate}

The tasks of preemptive toxicity detection are still in its infancy, and there is still a lot of room for research. For example, work so far has focused on using regular conversation features as predictors. Future work could look at adding toxic text features such as insults and curse words, or even using the output of single-line toxicity detection tools as features.

\section*{Acknowledgments}
This research was made possible by the financial, material, and technical support of Two Hat Security Research Corp., and the financial support of the Canadian research organization MITACS.

\bibliography{anthology,acl2020}
\bibliographystyle{IEEEtran}

\end{document}